\pdfoutput=1
\documentclass[a4paper]{PoS}
\usepackage{xspace} 
\usepackage{aas_macros} 
\usepackage[load-configurations = abbreviations,load=accepted]{siunitx} 
\DeclareSIUnit\ergon{erg}
\usepackage[numbers]{natbib}
\usepackage{subfig,graphicx}
\newcommand{\gammapy}{Gammapy\xspace}
\newcommand{\sherpa}{Sherpa\xspace}

\title{\gammapy: high level data analysis for extragalactic science
  cases with the Cherenkov Telescope Array}

\ShortTitle{\gammapy: extragalactic science cases with the Cherenkov Telescope Array}

\author{\speaker{J. Lefaucheur}$^{\,a}$, C. Boisson$^a$, Z. Bosnjak$^b$,
  M. Cerruti$^c$, C. Deil$^d$, J.~P. Lenain$^c$, S. Pita$^e$ and A. Zech$^a$
  for the CTA consortium\\
  \llap{$^a$}LUTH, Observatoire de Paris, PSL Research University, CNRS, Universit\'e Paris Diderot\\
  5 Place Jules Janssen, 92190 Meudon, France\\
  \llap{$^b$}Faculty of Electrical Engineering and Computing\\
  University of Zagreb, 10000 Zagreb, Croatia\\
  \llap{$^c$}Laboratoire de Physique Nucl\'eaire et de Hautes Energies (LPNHE)\\
  4 place Jussieu, F-75252, Paris Cedex 5, France\\
  \llap{$^d$}Max-Planck-Institut f\"ur Kernphysik\\
  P.O. Box 103980, D 69029 Heidelberg, Germany\\
  \llap{$^e$}APC, AstroParticule et Cosmologie, Universit\'{e} Paris Diderot, CNRS/IN2P3, CEA/Irfu,
  Observatoire de Paris, Sorbonne Paris Cit\'{e}\\
  10, rue Alice Domon et L\'{e}onie Duquet, 75205 Paris Cedex 13, France\\
  E-mail:  \email{julien.lefaucheur@obspm.fr}
}

\abstract{The Cherenkov Telescope Array (CTA) observatory will probe the non-thermal
  universe above 20 GeV up to several hundreds of TeV with a significant improvement
  in sensitivity and angular resolution compared to current experiments.
  Its outstanding capabilities will allow to increase the number of extragalactic
  cosmic accelerators detected at very high energy (VHE) and therefore to better
  constrain the population of VHE accelerators and the gamma-ray absorption processes
  in the intergalactic medium.
  For the first time in the history of imaging atmospheric Cherenkov telescopes (IACTs),
  CTA will be an open observatory and high-level data will be made available to the
  astronomical community.
  \gammapy is an open-source Python package developed by the Cherenkov telescope
  community that provides tools to simulate the gamma-ray sky and analyse IACT data.
  The versatile architecture of, and steady user contributions to \gammapy
  enable a large variety of high-level data analyses.
  Examples of \gammapy applications are presented, particularly in the context of
  extragalactic science with CTA.}

\FullConference{35th International Cosmic Ray Conference -ICRC217-\\
		10-20 July, 2017\\
		Bexco, Busan, Korea}

\begin{document}

\section{Introduction}
\label{sec:intro}
The next generation of Imaging Atmospheric Cherenkov Telescopes,
the Cherenkov Telescope Array,
will probe the $\gamma$-ray sky above \SI{20}{\GeV} with an
unmatched sensitivity and angular resolution using more than 100 telescopes
distributed on the South (Chile) and on the North (La Palma) sites.
To reach this goal, three different types of telescopes will be used~:
the \SI{23}{\meter} diameter Large-Size Telescopes (LST),
the \SI{12}{\meter} diameter Medium-Size Telescopes (MST) and
the \SI{4}{\meter} diameter Small-Size Telescopes (SST).
They are respectively adapted to detect photons of
\SIrange{20}{200}{\GeV} (field of fiew of \SI{\sim 4}{\degree}),
\SIrange{0.1}{10}{\TeV} (field of fiew of \SI{\sim 7}{\degree}) and
\SIrange{1}{300}{\TeV} (field of fiew of \SI{\sim 9}{\degree}).

With its improved sensitivity and higher field of view compared to current
experiments, namely MAGIC, H.E.S.S. and VERITAS, CTA will improve
our current knowledge on active galactic nuclei (AGN) and especially on blazars.
Nowadays, population studies are quite limited by the low/biased sample
of blazars detected at very high energy ($E \SI{\geq 100}{\GeV}$).
The high quality spectral resolution of CTA will help to look for
hadronic signatures in AGN spectra \citep{CTAHadr}.
Furthermore, it will also help to better constrain the intergalactic
medium opacity to $\gamma$-rays, and especially the flux density of the
diffuse extragalactic background light (EBL).
The first VHE detection of $\gamma$-ray bursts (GRBs) might shed light
on the radiative processes and the particle content in those accelerators
and might provide a new class of distant objects to constrain the EBL density.

CTA is expected to operate for thirty years and unlike current experiments
it will be an open observatory, meaning that high level data, such as event lists
and instrument response functions (IRFs), will be made available to the scientific
community.
\gammapy\footnote{\protect\href{http://docs.gammapy.org/}{http://docs.gammapy.org/en/latest/}}
is a community-developed, open-source Python package for
$\gamma$-ray astronomy \cite{CTA_GAMMAPY}.
It is built on widely-used scientific packages (Astropy, Numpy, Scipy)
and provides tools to simulate and analyse IACT data.
It was thought to produce high level data, such as spectra, maps, light curves
and even catalogues, taking as inputs lower level data such as event lists
and instrument response functions.
Classical methods to analyse data in very high energy astronomy, such as
background substraction or full-forward folding spectral reconstruction
are implemented and were highly tested with real data, especially in the H.E.S.S.
experiment.
Cube-style analysis, in which a spectral and a spatial model are jointly adjusted,
is also implemented \cite{LEA}.

In this contribution, we show how \gammapy can easily be  used to tackle
some extragalactic science cases of CTA.
We first present in Section \ref{sec:simu} how point-like source simulations
are made within \gammapy.
Section~\ref{sec:agn_pop} describes how it can contribute to the ongoing efforts
to estimate the number of expected blazars that could be detected with CTA.
Section~\ref{sec:ebl} is a technical discussion on how Gammapy can be used
to constrain the EBL scale factor. Section~\ref{sec:grb}
briefly presents an application about GRB detectability.
The Python scripts used to produce all results and figures presented here
are available at \protect\href{https://github.com/gammapy/icrc2017-gammapy-cta-egal}{https://github.com/gammapy/icrc2017-gammapy-cta-egal}.


\section{Simulations}
\label{sec:simu}
In the following, we will focus on point-like source simulations.
To reach this goal, simulations are generated with a set of instrument
response functions, which are produced by the simulation group in the consortium
\cite{CTA_IRF}, including~:
\begin{itemize}
\item effective area as function of true energy,
\item background rate as function of reconstructed energy,
\item energy migration matrix from true to reconstructed energy
\end{itemize}
An energy dependent-angular cut is applied to maximise the
significance in each interval of reconstructed energy.
To compute an expected excess for a given spectral models and exposure
we multiply the desired input spectra with the absorption of one of the available
EBL model \citep{ebl_dominguez,ebl_franceschini, ebl_finke} in \gammapy.
We sum the randomised expected excess and background counts to get the data
in the ON region of the source.
Finally, we randomised the background counts multiplied by the normalisation between
the ON and the OFF regions, taken as \SI{0.2}{} in the following,
to get the data in the OFF region.
The average time to run 100 simulations is less than \SI{900}{\milli \s} on a personal
computer\footnote{This estimation has been realised with the Timit Python package
  and a \SI{2.9}{\GHz} Intel Core i7 processor.}.
In the following sections, the level of significance is estimated with the
formula 17 of Li \& Ma \cite{lima}.

\section{Application example~: extrapolation of blazars from Fermi/LAT catalogues}
\label{sec:agn_pop}
The Fermi/LAT collaboration recently released the 3FHL (The Third Catalog of Hard
Fermi-LAT Sources) catalogue \citep{fermi3fhl}.
It contains \SI{1558} sources detected above \SI{10}{\GeV} in
84 months and analysed with the improved PASS8 analysis.
The blazars dominate the sample of sources (\SI{78}{\percent}),
followed by the galactic sources (\SI{9}{\percent}) and the unassociated
ones (\SI{13}{\percent}).
One of the main tasks of the extragalactic group of the CTA consortium,
in the preparatory phase,
is to estimate the number of detectable AGN by CTA, using as a starting point
the 3FHL catalogue, and to measure its impact on population studies.
A realistic analysis is available in \cite{CTA_AGNPOP}.

To illustrate \gammapy capabilities, we studied the impact
of adding an hypothetical exponential cut-off to the 3FHL Fermi/LAT spectra
on the detection level.
We selected a sample of 466 blazars, firmly identified or associated BL Lacs
or flat spectrum radio quasars (FSRQs), having a redshift estimation in the
catalogue and whose meridian transit occurs with zenith distance less
than \SI{30}{\degree} at least in one CTA site (north or south).
We applied an exponential cut-off at energy $\SI{1}{\TeV} / (1+z)$ to the
Fermi/LAT spectrum.
To assess the detection level of a source, the following procedure was used~:
\begin{itemize}
\item the spectrum is absorbed by the EBL model Dominguez et al. \cite{ebl_dominguez}
\item the IRF labelled as ``5h'', production 3b, are used according to the
  declination of the source ($\delta \geq 0$ is north site)
\item the simulation is done for \SI{20}{\hour} of observation time
\item the counts are integrated from \SI{50}{\GeV} to \SI{100}{\TeV}
\item the final significance is averaged on 20 simulations
\end{itemize}

\begin{figure}[t]
  \center 
  \includegraphics[width=.6\textwidth]{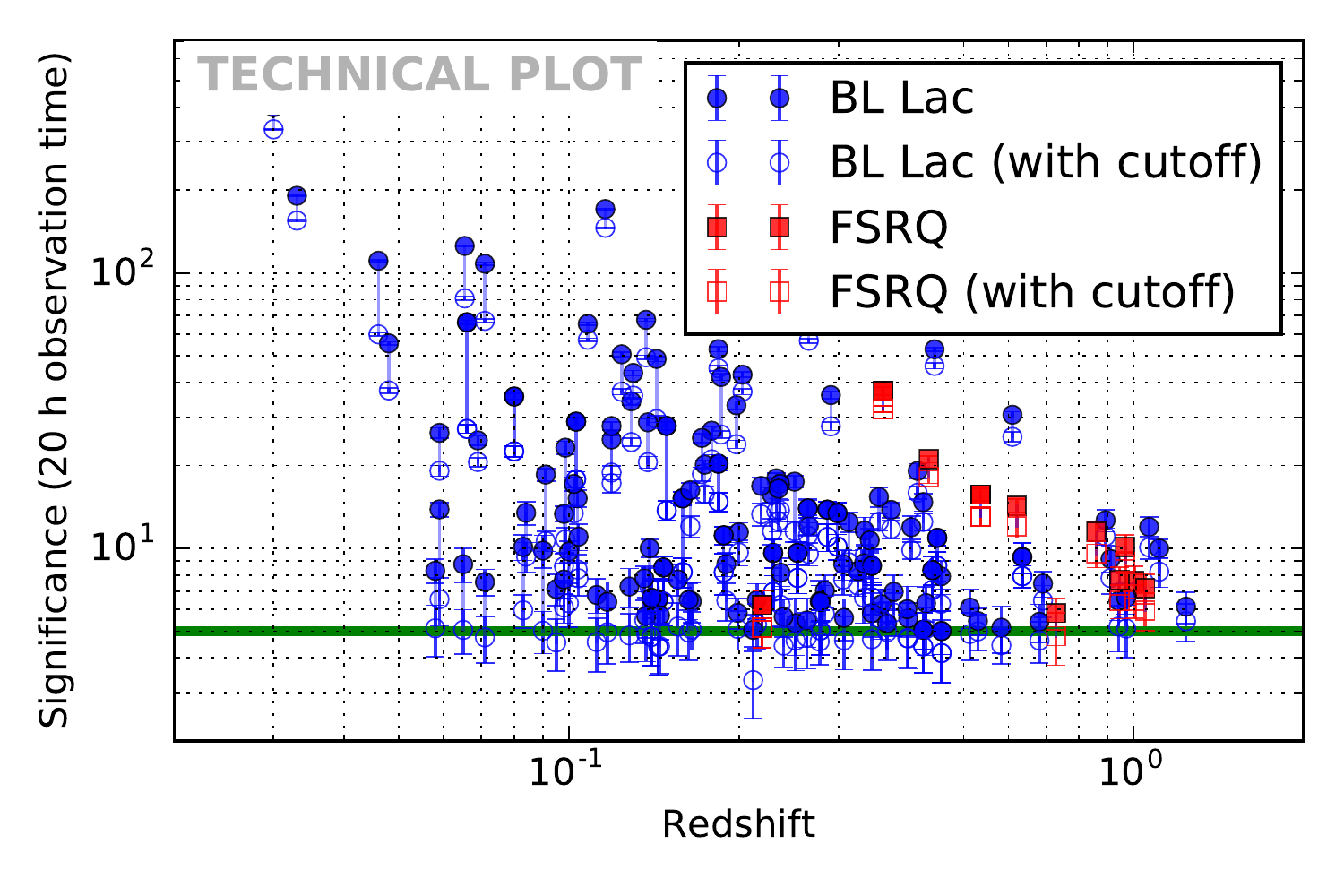}
  \caption{Average significance as a function of redshift for the simulated
    BL Lacs (blue circles) and FSRQs (red triangles). Full markers represent the
    3FHL spectral model and empty markers represent the 3FHL spectral model
    multiplied by an exponential cut-off at $\SI{1}{\TeV} / (1+z)$.}
  \label{fig:agn_pop}
\end{figure}
The results of the simulations are shown on Figure~\ref{fig:agn_pop}.
The fraction of sources detected above $5\sigma$ is \SI{\sim 53}{\percent} (151/290)
for BL Lacs and \SI{\sim 10}{\percent} (15/156) for FSRQs.
Adding the cut-off decreases theses numbers to \SI{46}{\percent}
and \SI{8}{\percent}, showing the impact of the choice of the intrinsic
spectral shape on detectability by CTA. 


\section{Application example: the EBL scale factor}
\label{sec:ebl}
One of the main cosmological topics with the Cherenkov
Telescope Array is to improve the constraints obtained with blazars on the
diffuse extragalactic background light.
A general and complete discussion can be found in \cite{CTA_EBL}.
In this section we will focus on the technical aspects to constrain an EBL scale factor
for a given absorption model with \gammapy and the Python
library \sherpa\footnote{\protect\href{http://cxc.cfa.harvard.edu/contrib/sherpa/}{http://cxc.cfa.harvard.edu/contrib/sherpa/}}.
\gammapy has been designed to provide flexibility in data-analysis
to the end users.
Simulations can be fed to widely-used tools that have been developed for
X-ray astronomy, such as the \sherpa library, to produce higher data level.
\sherpa gives access to the full-forward folding method
to handle spectral reconstruction and provides a convenient
way to do arithmetic with spectral models.
Furthermore, each model can be adjusted on its own dataset and each
parameter can be linked between different models, which makes
the fit of the EBL scale factor straightforward.
To obtain strong constraints on the EBL scale factor it is necessary to
adjust multiple sources for different redshift ranges \cite{CTA_EBL}.
The choice of the modelling of the source will have an impact on 
our ability to reconstruct the EBL scale factor.

As an illustration, we studied the impact of an exponential cut-off,
fixed at energy $\SI{1}{\TeV} / (1+z)$, 
on the determination of the EBL scale factor, called $\alpha$, for different redshifts.
To proceed, we used the same sample of blazars described in \cite{CTA_EBL}
and adjusted for each source the intrinsic spectra $\phi_{\text{int}}$ and $\alpha$
resulting in an observed spectrum~: 
$\phi(E)=\phi_{\text{int}}(E) \times e^{(-\alpha \tau_m(E,z))}$
where $\tau_{m}$ is the $\gamma$-ray opacity of the intergalactic medium
given by the model $m$.
For each source, the following procedure was used to study the EBL scale
reconstruction without (with) a cut-off~:
\begin{itemize}
\item the intrinsic spectrum is given by the 3FHL catalogue,
  e.g. a power law or a log-parabola (and multiplied by a cut-off)
\item the spectrum is absorbed by the EBL model \cite{ebl_dominguez}
\item the IRF labelled as ``5h'', production 3b, is chosen according to the
  declination of the source ($\delta \geq 0$ is north site)
\item the simulation is done for \SI{100}{\hour} of observation time
\item the spectrum is fitted with the full-forward folding method in
  a fixed energy range from \SI{50}{\GeV} to \SI{5}{\TeV},
  with the intrinsic spectra model (and multiplied by a cut-off)
\item the confidence interval bounds are computed for each parameter
\end{itemize}

\begin{figure}[t]
  \centering
  \subfloat[]{
    \label{fig:ebl}
    \includegraphics[width=0.48\columnwidth]{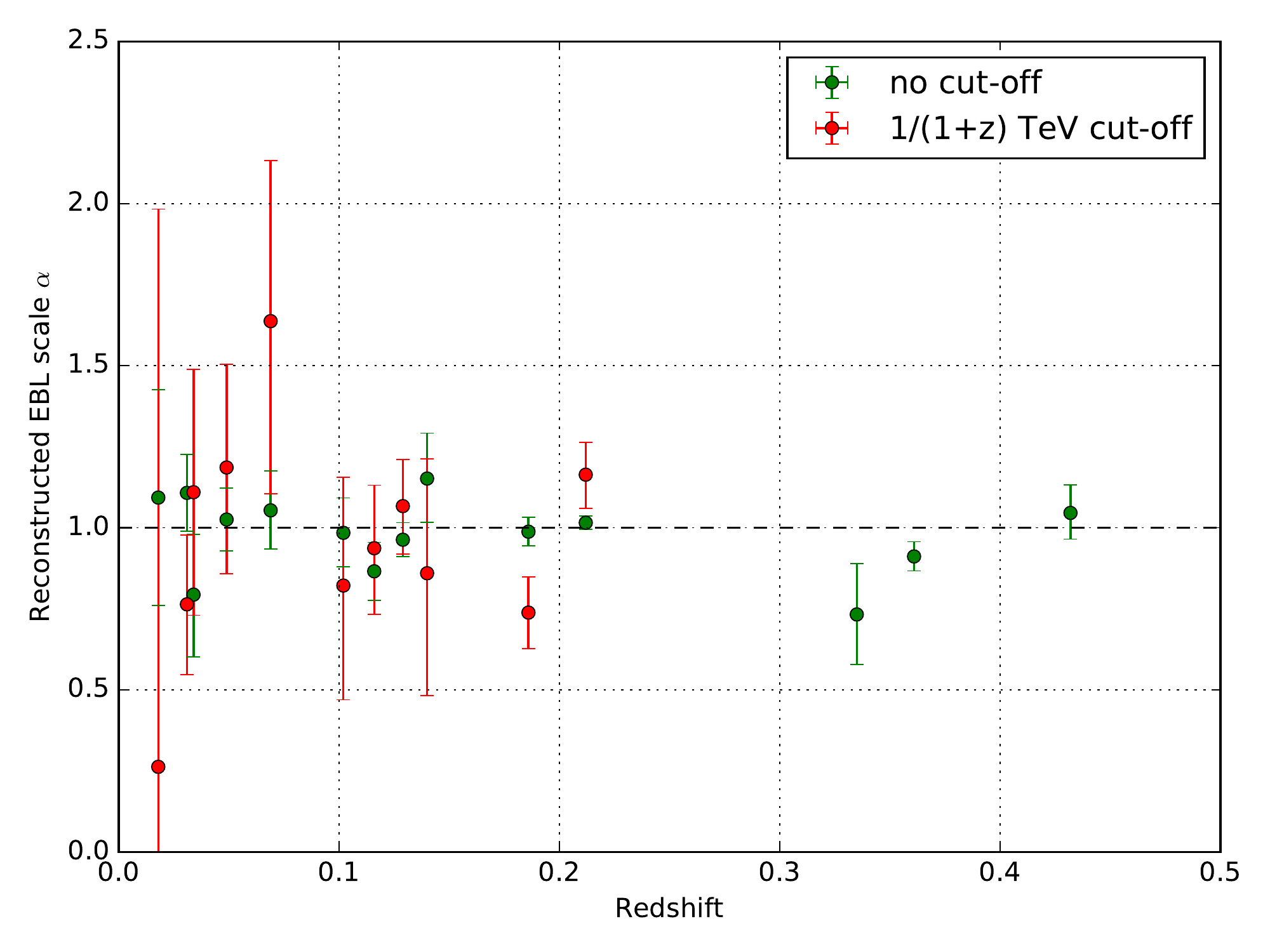}
  }
  \subfloat[]{
    \label{fig:attenuation}
    \includegraphics[width=0.48\columnwidth]{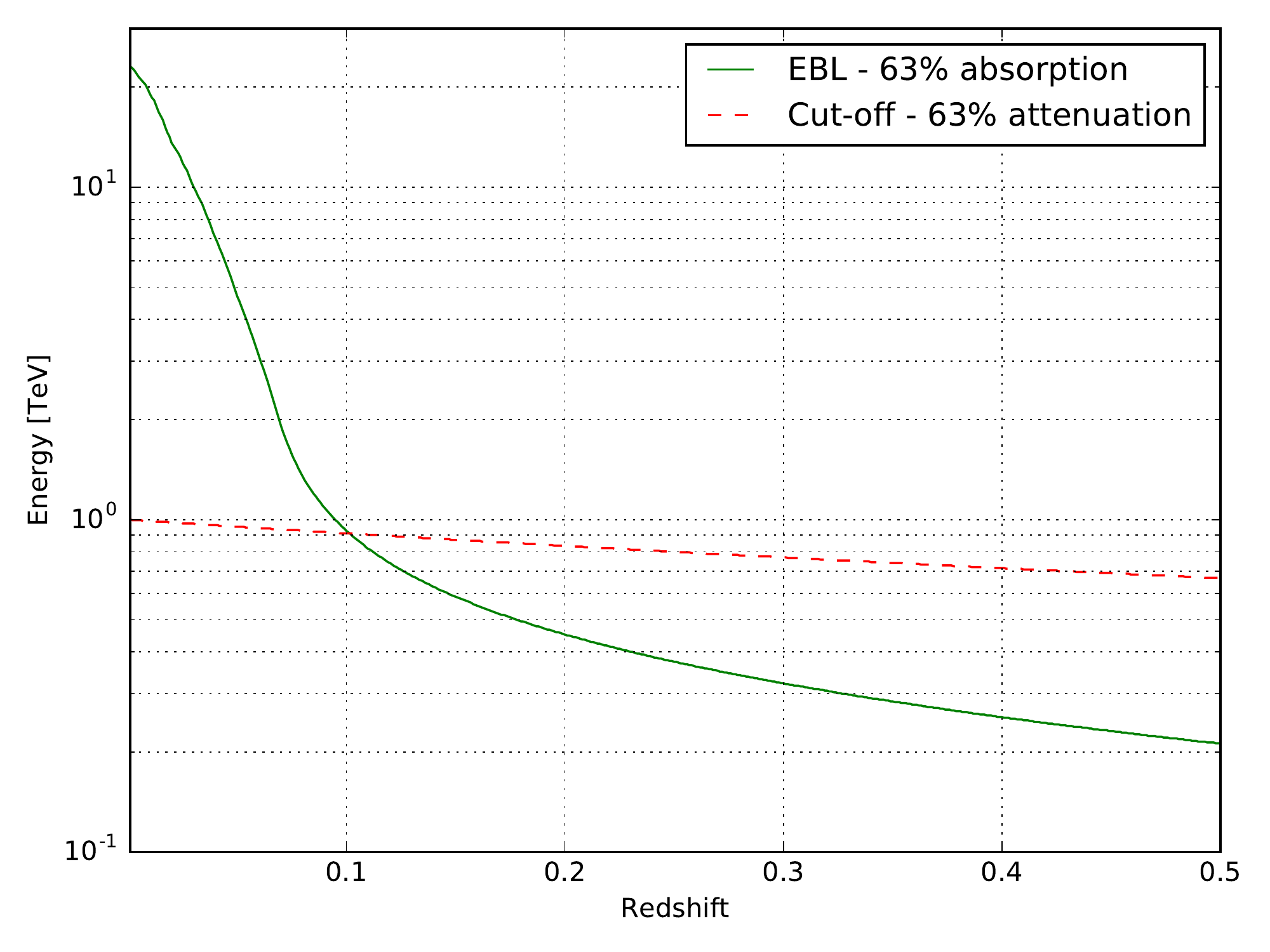}
  }\\ [-2ex]

  \caption{(a) Estimation of the reconstructed EBL scale $\alpha$ for
    individual sources used in \cite{CTA_EBL} as a function of their redshift.
    The best estimation of $\alpha$ obtained without and with an additional cut-off
    on the intrinsic spectral model are respectively shown with green and red points.
    (b) The curves, displayed as a function of energy and redshift,
    corresponding to a loss of \SI{63}{\percent} of the photons
    are respectively shown in green and red for the EBL absorption and the
    exponential cut-off attenuation.}
  \label{fig:ebl_blabla}
\end{figure}
We show on the technical Figure \ref{fig:ebl}, the reconstructed scale $\alpha$
as a function of the redshift of the sources for the two different spectral models.
For low redshift, $z \leq 0.1$, 
the EBL absorption begins in the energy domain where the exponential cut-off
is already present (see Figure~\ref{fig:attenuation}) so it makes it harder
to constrain the EBL scale factor (the error bars increase).
This could be in part compensated by a higher number of sources at low redshift.
For the intermediate redshift range, typically from $z \sim 0.1$ to $z \sim 0.2$,
the two effects are competitive and we are able to distinguish the two processes
involved in the photon attenuation. 
At higher redshifts, the EBL extinction intervenes well before the exponential cut-off
and the latter becomes less pertinent and thus under-constrained (for those cases, the maximisation
likelihood procedure fails).
That shows that \gammapy, in connection with the \sherpa library,
is particularly well suited for these kinds of studies involving custom
complex spectral models.

\section{Application example: observational window for $\gamma$-ray bursts}
\label{sec:grb}
$\gamma$-ray bursts are transient and explosive phenomena
originating from cosmological distances. For the brief episode of emission
their total energy output
reaches 10$^{51}$-10$^{53}$ ergs, placing them among the most powerful
presently observed objects.
A burst is characterised by a prompt phase dominated by X-ray and
$\gamma$-ray photons (typically between 10 keV
and a few MeV), lasting for a few milli-seconds to a few hundreds of seconds.
It is followed by a second emission phase, called the afterglow, during
which the electromagnetic emission is shifted to lower energies (X-rays, visible,
radio) and rapidly decaying with time.
The origin of the bursts is not fully understood. Nevertheless, the
combination of the short variability timescale and the huge energy release
suggest that GRBs are the results of cataclysmic events in the Universe,
most likely associated with the births of stellar size black holes or
rapidly spinning, highly magnetized, neutron stars.
Up to now, the Fermi/LAT detected more than 100 $\gamma$-ray bursts above
\SI{75}{\MeV},
whereas IACTs never detected a significant $\gamma$-ray emission from a GRB.
With its high effective area and its big field of view, CTA might detect a
few of these events and help to better constrain their intrinsic properties.
To get a more detailed picture about high energy and very high energy GRBs
the reader may refer to \cite{Piron:2016aa} and \cite{Inoue:2013aa},
respectively.


Here, we used \gammapy to reproduce the result from \cite{CTA_GRB} showing
the detection significance of the bright $\gamma$-ray burst GRB~080916C as a function of the
observational window.
The spectral template is described
as a power law of spectral index of $2$ with an integrated
flux in the \SIrange{0.1}{10}{\GeV} range at a time $t_p$ of
$\Phi (t_p=\SI{6.5}{\s}) = \SI{500e-5}{\per \square \cm \per \s}$.
We further assume that the GRB is located at a distance\footnote{The
  measured redshift of GRB~080916C is $z=4.35 \pm 0.15$ \cite{GRB0809}.}
corresponding to $z=3$.
The flux decay over time is parameterized by a decay index
$\delta=\SI{1.7}{}$.
In order to study the significance of the source as a function of the
observational window, we defined 20 logarithmic time intervals representing the beginning
and the end of the observations, from \SI{20}{\s} to 2 days.
Starting from $t_0 = \SI{20}{\s}$ after the peak flux, the significance was computed for
each physical logarithmic time interval in the following way~:
\begin{itemize}
\item the average time interval $t_{\text{int}}$ is computed by taking into account the flux decay law
\item the flux normalisation is computed for $t_{\text{int}}$
\item the spectrum is absorbed with the EBL model from \cite{ebl_dominguez}
\item the North IRF, labelled as ``0.5h'' from the production 2 are used
\item the counts are integrated from \SI{30}{\GeV} to \SI{1}{\TeV}
\end{itemize}

\begin{figure}[t]
  \center 
  \includegraphics[width=.6\textwidth]{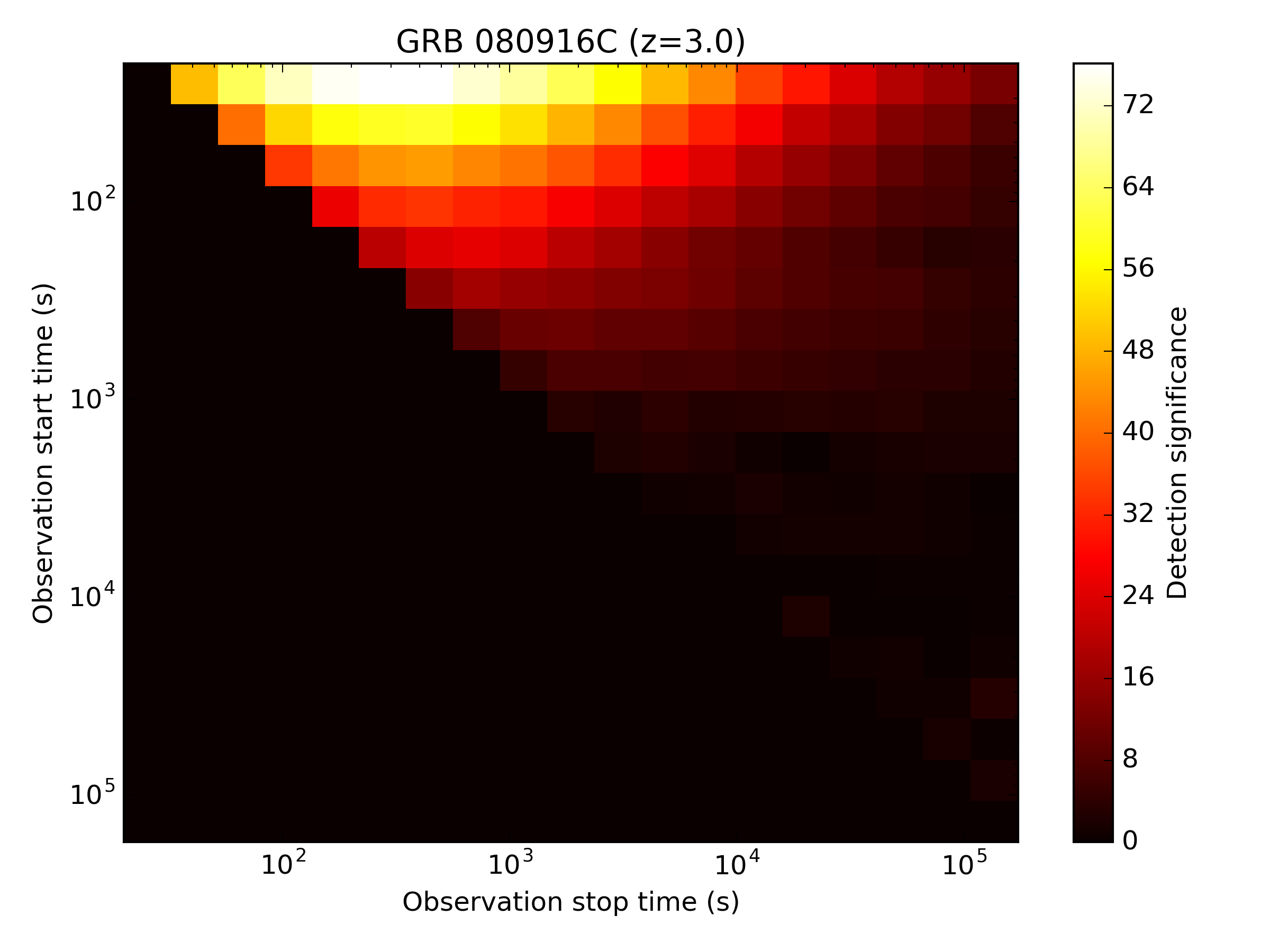}
  \caption{Significance of detection as a function of time windows (t-start, t-stop)
    for the template of GRB as described in the text.}
  \label{fig:grb_twindow}
\end{figure}
Results of the simulations are shown on Figure~\ref{fig:grb_twindow}.
The variation of the significance
results from two effects, the flux decay of the source and a constant level of background as a function of time.

\section{Conclusion}
\label{sec:conclusion}
\sloppy
We have shown how \gammapy capabilities can be used to 
handle extragalactic science cases with the Cherenkov Telescopes Array.
As mentioned in the introduction, we note that the Python scripts used to
produce all results and figures presented in this contribution are available
at \protect\href{https://github.com/gammapy/icrc2017-gammapy-cta-egal}{https://github.com/gammapy/icrc2017-gammapy-cta-egal}. 
This means anyone can reproduce and check the results, and use the scripts
as examples to start their own studies.

\acknowledgments
This work was conducted in the context of the CTA Consortium.
Julien Lefaucheur would like to thank R\'egis Terrier (APC/CNRS) for useful
discussions about data analysis with the Python library \sherpa.
\bibliographystyle{JHEP}
\bibliography{biblio}



\end{document}